\documentclass{article}%
\usepackage{amsmath}
\usepackage{graphicx}
\usepackage{wrapfig}%
\usepackage{amsfonts}%
\usepackage{amssymb}
\begin{document}
\textbf{Specific Heat of Ce}$_{1-x}$\textbf{La}$_{x}$\textbf{RhIn}$_{5}$
\textbf{in Zero and Applied Magnetic Field: \ A Very Rich Phase Diagram}

\bigskip

J. S. Kim, J. Alwood, D. Mixson, P. Watts, and G. R. Stewart

Department of Physics, University of Florida, Gainesville, Fl. \ 32611-8440

\bigskip

Abstract: \ Specific heat and magnetization results as a function of field on
single- and poly-crystalline samples of Ce$_{1-x}$La$_{x}$RhIn$_{5}$ show 1.)
a specific heat $\gamma$ of about 100 mJ/moleK$^{2}$ (in agreement with recent
dHvA results of Alvers et al.); 2.) upturns at low temperatures in C/T and
$\chi$\ that fit a power law behavior (%
%TCIMACRO{\TEXTsymbol{<}}%
%BeginExpansion
$<$%
%EndExpansion
=%
%TCIMACRO{\TEXTsymbol{>} }%
%BeginExpansion
$>$
%EndExpansion
Griffiths phase non-Fermi liquid behavior); 3.) a field induced anomaly in C/T
as well as M vs H behavior in good agreement with the recent Griffiths phase
theory of Castro Neto and Jones, where M\symbol{126}H at low field,
M\symbol{126}H$^{\lambda}$ above a crossover field, C/T\symbol{126}%
T$^{-1+\lambda}$ at low field, and C/T\symbol{126}(H$^{2+\lambda/2}%
$/T$^{3-\lambda/2}$)*exp(-$\mu_{eff}$H/T) above the same crossover field as
determined in the magnetization and where $\lambda$ is independently
determined from the temperature dependence of $\chi$ at low temperatures,
$\chi$\symbol{126}T$^{-1+\lambda}$ and low fields.

\pagebreak 

\textbf{I \ Introduction}

\bigskip

\qquad Recently, a new family of heavy-fermion compounds has been discovered
that crystallize in a layered, tetragonal structure with chemical composition
CeMIn$_{5}$, where M= Ir, Co, and Rh. Characteristic of heavy-fermion systems,
each member exhibits a large Sommerfeld coefficient $\gamma$ ($\equiv$C/T as
T$\rightarrow$0) in the specific heat C. CeIrIn$_{5}$ and CeCoIn$_{5}$ are
bulk superconductors$^{1-2}$ with transition temperatures at T$_{c}$ = 0.4 K
and 2.3 K and normal-state values of $\gamma\approx$ 750 mJ/molK$^{2}$ and
1200 mJ/molK$^{2}$, respectively. CeRhIn$_{5} $ displays heavy-fermion
antiferromagnetism with$^{3}$ T$_{N}$ = 3.8 K. A precise value of $\gamma$ is
difficult to establish unambiguously because of the N\'{e}el order; a lower
limit of approximately 400 mJ/molK$^{2}$ has been quoted$^{4-5}$.

\qquad In our high field specific heat measurements$^{6}$ on the CeMIn$_{5}$
compounds, we found that the large upturn for M=Rh in C/T above T$_{N}$ (C/T
is already 1000 mJ/molK$^{2}$ at T$_{N}$) as temperature is lowered appeared
to be primarily due to magnetic interactions above the antiferromagnetic
transition since the specific heat data at a given temperature for T%
%TCIMACRO{\TEXTsymbol{>}}%
%BeginExpansion
$>$%
%EndExpansion
T$_{N}$ in different fields up to 32 T all coincide with one another when the
temperature axis was scaled to T/T$_{N}$. \ Recently Alver, et al. have
performed$^{7}$ dHvA measurements on twelve single crystal samples spanning
the whole composition range of Ce$_{1-x}$La$_{x}$RhIn$_{5}$ and find rather
low (i. e. inconsistent by approximately an order of magnitude with a $\gamma$
of 400 mJ/molK$^{2})$ effective masses from the dilute Ce, large x end of the
phase diagram up to x=0.1. \ At this Ce-rich end of the composition range they
find an increase in the effective masses (which still remain $\leq$ 10 m$_{e}%
$) which they ascribe to spin fluctuation effects. \ Alver, et al. conclude
that the Ce f-electrons remain localized in Ce$_{1-x}$La$_{x}$RhIn$_{5}$ for
all x, with the (modest) observed mass enhancement near pure CeRhIn$_{5}$ due
to spin fluctuation effects. \ Although comparisons between specific heat and
dHvA data have inherent problems (not the least of which is the possibility of
unseen, heavier mass orbits in the dHvA measurements), an effective mass
enhancement of approximately ten normally corresponds to a specific heat
$\gamma$ of only \symbol{126}50 mJ/molK$^{2}$. \ This is a wide discrepancy
from the estimate of 400 mJ/molK$^{2}$ in$^{4-5}$ the literature; this
discrepancy would be consistent with our high field specific heat result$^{6}$
that the upturn above T$_{N}$ in C/T in pure CeRhIn$_{5}$ is primarily caused
by magnetic interactions, which would not cause a mass enhancement observable,
e.g., in dHvA measurements.

\qquad In order to help resolve this seeming disagreement, to determine the
specific heat $\gamma$ (also proportional to the effective mass) in a region
of the phase diagram away from the antiferromagnetic anomaly, and to look for
possible new behavior in the dilute limit we report here on a specific heat
study of both single and polycrystalline samples of Ce$_{1-x}$La$_{x}%
$RhIn$_{5}$, 0$\leq$x$\leq$0.95. \ Certainly, doping studies$^{8-10}$ on other
heavy Fermion systems, e.g. Ce$_{1-x}$La$_{x}$Cu$_{2}$Si$_{2}$, Ce$_{1-x}%
$Th$_{x}$Cu$_{2}$Si$_{2}$, and U$_{1-x}$Th$_{x}$Be$_{13}$, have revealed
interesting new information - both about the respective parent compound as
well as new physics in the dilute limit. \ Polycrystalline samples were
originally chosen for the study as being more easily and rapidly prepared.
\ However, specific heat results for polycrystalline Ce$_{1-x}$La$_{x}%
$RhIn$_{5}$, x=0.5 and 0.8 were determined to disagree with specific heat
results for single crystal samples, while results agreed for x=0.15 and 0.95.
\ This disagreement appears due to the presence of a second phase which we
were able to eliminate through long term annealing of the polycrystalline
samples at a relatively low temperature.

\bigskip

\textbf{II \ Experimental}

\bigskip

\qquad Single crystal samples of Ce$_{1-x}$La$_{x}$RhIn$_{5}$ were prepared
using the procedure described in ref. 6, which was similar to that used in
refs. 4 and 7. \ Excess In was removed from the resulting flat platelet
crystals using an H$_{2}$O:HF:H$_{2}$O$_{2}$ 4:1:1 etch which was different
than the centrifugal method (H$_{2}$O:HCl 4:1 etch) used in ref. 4\ (7);
however the present work's specific heat results (which are a measure of bulk
properties) should be relatively independent of such surface treatments. \ The
polycrystalline samples in the present work (previous work in the literature
has been almost uniformly on single crystal samples) were prepared by melting
together stoichiometric amounts of the appropriate high purity starting
elements (using Ames Laboratory Ce and La, 99.95\% pure Rh from Johnson Mathey
Aesar, and 99.9999\% In from Johnson Matthey Aesar - the same starting
materials as used for the single crystals) under a purified inert Ar
atmosphere. \ Weight losses after four melts, with a flipping of the
arc-melted button between melts to improve homogeneity, were in the range of
1\%, primarily due to In loss. \ Additional In was added in the beginning to
correct for this, such that the In concentrations after the last melt were
within $\pm$0.2\% of the stoichiometric amount.

\ \qquad Specific heat in fields to 13 T were measured using established
techniques$^{11}$, while magnetic susceptibility data were measured in a SQUID
magnetometer from Quantum Design. \ 

\bigskip

\textbf{III Results and Discussion}

\bigskip

\qquad Figure 1 shows the specific heat divided by temperature vs temperature
for single crystal Ce$_{1-x}$La$_{x}$RhIn$_{5}$, x=0, 0.15, 0.5, 0.8, and 0.95
and polycrystal Ce$_{1-x}$La$_{x}$RhIn$_{5}$, x=0.32. \ All samples were
single phase. \ Results for unannealed polycrystalline Ce$_{1-x}$La$_{x}%
$RhIn$_{5}$, x=0.15 and 0.95, and annealed (35 days at 720 $^{o}$C)
polycrystalline Ce$_{1-x}$La$_{x}$RhIn$_{5}$, x=0.5 and 0.8, were comparable
to the single crystal results (see inset of Fig. 1 for an example); however,
unannealed polycrystalline samples for x=0.5 and 0.8 contained a second phase
that ordered antiferromagnetically below 1 K. \ This was taken as a sign of an
incipient miscibility gap which - due to previous work being focussed on
single crystal samples - was heretofore unknown. \ 

\qquad From the data shown in Fig. 1, one can follow the suppression of the
antiferromagnetic transition with increasing La doping; \ there is a clear,
although reduced in magnitude, transition at 2 K for 15\% La doping that is
absent by x=0.32. \ Although one might expect$^{12}$ non-Fermi liquid ('nFl')
behavior when \bigskip T$_{N}$ is suppressed to T=0, the temperature
dependence of the C/T data for x=0.32 - although the

\bigskip%
%TCIMACRO{\FRAME{ftbpFU}{3.2897in}{2.3497in}{0pt}{\Qcb{C/T vs T for
%Ce(1-x)La(x)RhIn(5)}}{}{lafig1js7.eps}{\special{ language "Scientific Word";
%type "GRAPHIC";  maintain-aspect-ratio TRUE;  display "USEDEF";
%valid_file "F";  width 3.2897in;  height 2.3497in;  depth 0pt;
%original-width 3.6547in;  original-height 3.1073in;  cropleft "0";
%croptop "0.9276";  cropright "1";  cropbottom "0.0904";
%filename 'LAFIG1JS7.EPS';file-properties "XNPEU";}} }%
%BeginExpansion
\begin{figure}
[ptb]
\begin{center}
\includegraphics[
trim=0.000000in 0.280900in 0.000000in 0.224968in,
natheight=3.107300in,
natwidth=3.654700in,
height=2.3497in,
width=3.2897in
]%
{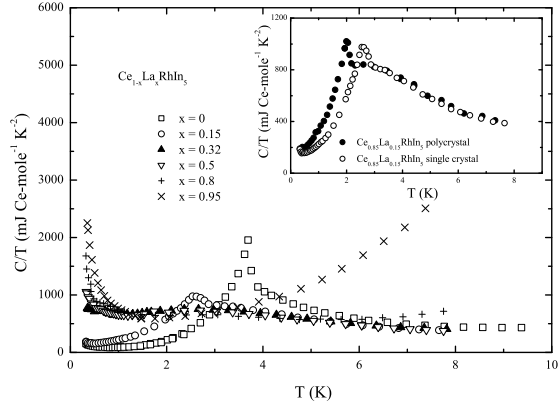}%
\caption{C/T vs T for Ce(1-x)La(x)RhIn(5)}%
\end{center}
\end{figure}
%EndExpansion

data show an upturn - is only measured for \symbol{126}0.5 K below the hump.
\ This is too restricted a temperature range to allow conclusions about the
temperature dependence.

\qquad Before we discuss the behavior of $\gamma$ as a function of x in
Ce$_{1-x}$La$_{x}$RhIn$_{5}$, we will first focus on the upturn at low
temperatures for x$\geq$0.5.

\bigskip

\textbf{A \ Upturn in C/T for x}$\geq$\textbf{0.5}

\qquad The upturn in C/T for x$\geq$0.5 in Ce$_{1-x}$La$_{x}$RhIn$_{5}$ shown
in Figs. 1 is fit in Figs. 2 and 3 for single crystalline, as well as single
phase polycrystalline, material. \ Note in Fig. 2 that the data for the three
different samples agree rather well. \ There is certainly no sign in the dHvA
results of Alver, et al. for a strong, heavy fermion upturn in C/T that would
cause large effective masses. \ Thus, this upturn at low temperatures in C/T
likely has a magnetic interaction explanation (see section \textbf{C} below
for the field dependence). \ The temperature dependence of the upturns in C/T
(see Figs. 2 and 3) for single crystal Ce$_{1-x}$La$_{x}$RhIn$_{5}$, x=0.5,
0.8, and 0.95, is not at all like the high temperature side of a Schottky peak
(C \symbol{126}1/T$^{2}$) but rather appears (in the somewhat limited
temperature range that we have data) to follow C/T \symbol{126}T$^{-1+\lambda
}$,$\lambda_{\text{C/T}}$=0.63 $\pm$ 0.1, 0.37 $\pm$ 0.1, and \symbol{126}0
respectively. \ This is the temperature dependence predicted for non-Fermi
liquid behavior caused by disorder-induced spin clusters, the so-called
Griffiths phase$^{12-13}$. \ (Note that the fits of $\chi$ to T$^{-1+\lambda}$
below 1.2 K are much better than fits to either log T or T$^{0.5}$.) \ In this
theory, the magnetic susceptibility%

%TCIMACRO{\FRAME{ftbpFU}{3.2897in}{2.2572in}{0pt}{\Qcb{C/T for 3 samples of
%x=0.5}}{}{lafig2js7.eps}{\special{ language "Scientific Word";
%type "GRAPHIC";  maintain-aspect-ratio TRUE;  display "USEDEF";
%valid_file "F";  width 3.2897in;  height 2.2572in;  depth 0pt;
%original-width 3.5968in;  original-height 3.0104in;  cropleft "0";
%croptop "0.8897";  cropright "1";  cropbottom "0.0735";
%filename 'LAFIG2JS7.EPS';file-properties "XNPEU";}} }%
%BeginExpansion
\begin{figure}
[ptb]
\begin{center}
\includegraphics[
trim=0.000000in 0.221264in 0.000000in 0.332047in,
natheight=3.010400in,
natwidth=3.596800in,
height=2.2572in,
width=3.2897in
]%
{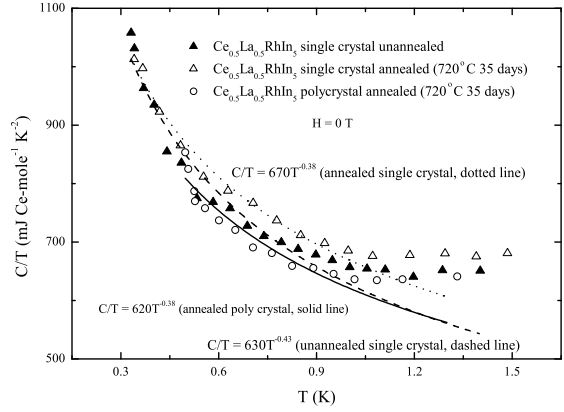}%
\caption{C/T for 3 samples of x=0.5}%
\end{center}
\end{figure}
%EndExpansion

\bigskip%
%TCIMACRO{\FRAME{ftbpFU}{3.2897in}{2.3506in}{0pt}{\Qcb{C/T vs T,
%x=0.5,0.8,0.95, fit to T\symbol{94}(-1+lambda)}}{}{lafig3js7.eps}%
%{\special{ language "Scientific Word";  type "GRAPHIC";
%maintain-aspect-ratio TRUE;  display "USEDEF";  valid_file "F";
%width 3.2897in;  height 2.3506in;  depth 0pt;  original-width 3.6547in;
%original-height 2.9404in;  cropleft "0";  croptop "0.9235";  cropright "1";
%cropbottom "0.0382";  filename 'LAFIG3JS7.EPS';file-properties "XNPEU";}} }%
%BeginExpansion
\begin{figure}
[ptb]
\begin{center}
\includegraphics[
trim=0.000000in 0.112323in 0.000000in 0.224941in,
natheight=2.940400in,
natwidth=3.654700in,
height=2.3506in,
width=3.2897in
]%
{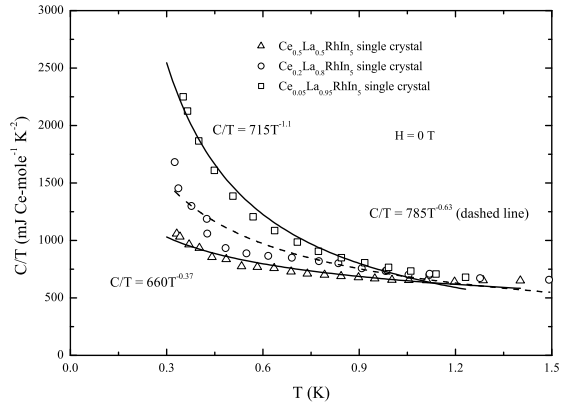}%
\caption{C/T vs T, x=0.5,0.8,0.95, fit to T\symbol{94}(-1+lambda)}%
\end{center}
\end{figure}
%EndExpansion

at low temperature should have the same power law dependence as C/T. \ The
susceptibility at low temperatures for these same compositions of single
crystal Ce$_{1-x}$La$_{x}$RhIn$_{5}$, see Fig. 4, does indeed fit this
T$^{-1+\lambda}$ temperature dependence, with $\lambda_{\chi}$ =\{0.73,0.90\},
\{0.50,0.70\}, \{0.14,0.30\} respectively for H \{$\perp$,$\parallel$\} the
c-axis, where the absolute error bar for each value is $\pm$ 0.1 (with,
however, somewhat better precision, useful for intercomparison between values
derived from a \textit{given} measurement technique. \ For example, 0.14
derived from $\chi$ for x=0.95\ is certainly less than 0.30 derived for the
other field direction, but is comparable to the value of \symbol{126}0 derived
for the same composition from the specific heat.) \ (Note that other standard
non-Fermi liquid temperature dependences, such as $\chi$ \symbol{126}log T or
T$^{0.5}$, do not fit the $\chi$ data at all well.) \ Although for a given
composition the respective exponents for C/T and $\chi$ agree within
experiment accuracy only for $\chi$(H $\perp$ c), the recent theory$^{14}$ of
Castro Neto and Jones actually predicts that $\chi$ and C/T may diverge
\textit{differently} at low temperature, relaxing the requirement of the early
theory$^{12-13}$ that $\lambda_{\chi}$=$\lambda_{\text{C/T}}$. It is clear
that the disorder requirement for uncompensated spins (which requires that M
vs H is shows saturation behavior) is fulfilled for all these compositions
(see discussion and accompanying figures in section \textbf{C} below.) \ In
addition, the agreement in $\lambda_{C/T}$ and $\lambda_{\chi}$found for the
upturn in C/T and $\chi$ in the present work is comparable to that found by,
e. g., deAndrade et al.$^{15}$ in their study of Th$_{1-x}$U$_{x}$Pd$_{2}%
$Al$_{3}$ - even though they measured $\chi$ down to 0.5 K, i. e. in a
temperature range comparable to that for their specific heat measurements.
\ \ The anisotropy of the susceptibility-determined $\lambda$ values is
thought to be real, and not related to the discrepancy between $\lambda
_{C/T\text{ }}$and $\lambda_{\chi}$.%

%TCIMACRO{\FRAME{ftbpFU}{3.2897in}{2.3575in}{0pt}{\Qcb{Chi vs T fit to
%T\symbol{94}(-1+lambda)}}{}{lafig4js7.eps}%
%{\special{ language "Scientific Word";  type "GRAPHIC";
%maintain-aspect-ratio TRUE;  display "USEDEF";  valid_file "F";
%width 3.2897in;  height 2.3575in;  depth 0pt;  original-width 3.6115in;
%original-height 2.9127in;  cropleft "0";  croptop "0.9237";  cropright "1";
%cropbottom "0.0381";  filename 'LAFIG4JS7.EPS';file-properties "XNPEU";}} }%
%BeginExpansion
\begin{figure}
[ptb]
\begin{center}
\includegraphics[
trim=0.000000in 0.110974in 0.000000in 0.222239in,
natheight=2.912700in,
natwidth=3.611500in,
height=2.3575in,
width=3.2897in
]%
{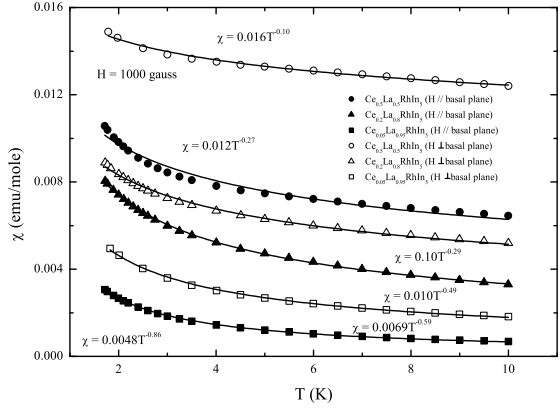}%
\caption{Chi vs T fit to T\symbol{94}(-1+lambda)}%
\end{center}
\end{figure}
%EndExpansion

\qquad As one possible check for a tendency towards magnetic behavior, the
Wilson ratio (R $\propto$ $\chi$/$\gamma\mu_{eff}^{2}$) - which is used$^{16}$
in the study of heavy Fermion systems to track the tendency towards magnetism,
with R $\succsim$ 0.8 indicating$^{16}$ magnetic behavior - for these
Ce$_{1-x}$La$_{x}$RhIn$_{5}$ alloys is in the range of 1.0 to 1.8, i. e. they
definitely show magnetic character. \ As a further check for evidence for spin
clusters, we investigated these compositions for spin glass behavior and - to
within the limits ($\pm$2 \%) of the accuracy of the measurements - found no
difference between field cooled and zero field cooled data down to 1.8 K.
\ This lack of observable spin glass behavior in the dc magnetic
susceptibility in these samples does not rule out a Griffiths phase
interpretation$^{17}$.

\textbf{B \ Specific Heat }$\gamma$ \textbf{as a Function of x}

\qquad The original goal of this work, besides the hope for new physics of
interest in the dilute range (already partially fulfilled by the results
discussed above for the low temperature upturn in C/T and $\chi)$ was to
investigate the specific heat $\gamma$ (defined as C/T as T$\rightarrow$0)
away from the region of the phase diagram where antiferromagnetism obscures
C/T as T$\longrightarrow$0 in CeRhIn$_{5}$ diluted with La. \ As discussed
above, after the antiferromagnetism is suppressed (x%
%TCIMACRO{\TEXTsymbol{>}}%
%BeginExpansion
$>$%
%EndExpansion
0.15), a low temperature upturn in the C/T data (Fig. 1) occurs that,
normalized per Ce-mole, becomes more pronounced with increasing dilution of
the Ce. \ This upturn appears not to be related to the effective masses
measured by the dHvA measurements. \ 

\qquad A further complication to determining the specific heat $\gamma$ is the
rounded feature in C/T centered at \symbol{126}3 K visible already for x=0.15
above T$_{N}$. \ As may be seen from Fig. 5, the C/T data for x$=$0.5
(triangles) and 0.8 (circles) in Ce$_{1-x}$La$_{x}$RhIn$_{5}$ above the low
temperature upturn show a tendency to curve or bend downwards down to about
1.5 K, at which point the upturn discussed in the section above begins. \ This
'hump' in C/T centered at \symbol{126}3 K makes extrapolating C/T to T=0 to
determine $\gamma$ a somewhat imprecise procedure. \ It should be stressed
that this rounded feature, or hump, in C/T has its provenance in the
f-electron sublattice: \ such a feature is \textit{not} present in C/T data
for pure LaRhIn$_{5}^{18}$. \ One possibility for correcting for this feature
in order to determine $\gamma$\ is to subtract off both the low temperature
upturn (see Fig. 3 for the fits to the upturns) \textit{and} a fit$^{18}$ to
pure LaRhIn$_{5}$ and examine the remainder. \ As shown in the inset to Fig. 5
for x=0.5, this very rough approximation (the apparent negative value below
about 1 K is, see Fig. 3, merely a sign that the fit to the upturn - which
goes up to over 1000 mJ/Ce-moleK$^{2}$ at 0.3 K - is in error as
T$\rightarrow$1 K) allows us to assign an approximate$^{19}$ $\gamma$ value
per Ce mole of $\leq$100 mJ/CemolK$^{2}$ for x$\geq$0.5. \ This agrees much
better with Alver, et al.'s dHvA results than the estimates of 400
mJ/CemolK$^{2}$ estimated$^{4-5}$ in the literature. \ However, as the La
dilution is removed, for x$\leq$0.1, Alver, et al. report approximately a
factor of two increase in effective mass due to spin fluctuation effects, with
an effective mass for pure CeRhIn$_{5}$ that would correspond to a $\gamma$ of
approximately 50 mJ/CemolK$^{2}$. \ In the dilute limit, Alver et al.'s
effective measured effective mass corresponds to a $\gamma$ of only 25
mJ/CemolK$^{2}$. \ However, as may be seen in Fig. 5, our C/T data at low
temperature are much too obscured by the unexpected upturn as well as\ by the
rounded maximum to supply any sort of accurate estimate for $\gamma$ beyond
the dilute, x$\geq$0.5, range of $\leq$100 mJ/CemolK$^{2}$ already quoted above.%

%TCIMACRO{\FRAME{ftbpFU}{3.2897in}{2.3938in}{0pt}{\Qcb{C/T vs T showing the
%'hump' at 3 K}}{}{lafig5js7.eps}{\special{ language "Scientific Word";
%type "GRAPHIC";  maintain-aspect-ratio TRUE;  display "USEDEF";
%valid_file "F";  width 3.2897in;  height 2.3938in;  depth 0pt;
%original-width 3.5968in;  original-height 2.9404in;  cropleft "0";
%croptop "0.9247";  cropright "1";  cropbottom "0.0376";
%filename 'LAFIG5JS7.EPS';file-properties "XNPEU";}} }%
%BeginExpansion
\begin{figure}
[ptb]
\begin{center}
\includegraphics[
trim=0.000000in 0.110559in 0.000000in 0.221412in,
natheight=2.940400in,
natwidth=3.596800in,
height=2.3938in,
width=3.2897in
]%
{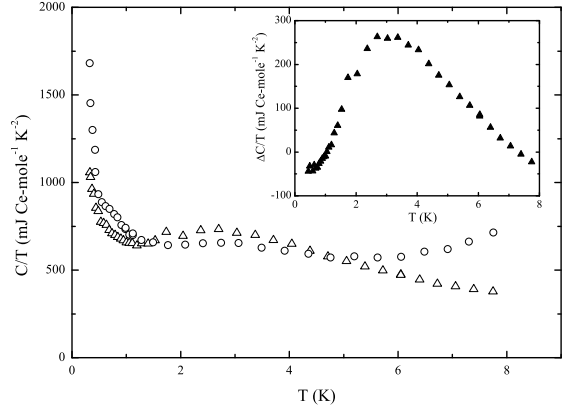}%
\caption{C/T vs T showing the 'hump' at 3 K}%
\end{center}
\end{figure}
%EndExpansion

\bigskip\textbf{C \ Field Induced Anomaly for x}$\geq$\textbf{0.5}

\qquad As a final aspect of new, unexpected behavior for CeRhIn$_{5}$ diluted
with La, when we were investigating the field dependence of the upturn in the
specific heat divided by temperature using magnetic field as a probe, we
discovered that applied field suppresses the low temperature upturn in C/T at
rather low field and induces an peak in C/T that, with increasing field, moves
up in temperature and becomes broader and less pronounced. \ This rounded
anomaly, shown in Fig. 6 for x=0.95 (these data are typical of the results for
all x$\geq$0.5) with field in the basal plane (data in the perpendicular
direction are within 15 percent of these), is not that of either a spin glass
(where C\symbol{126}1/T above the peak) or a Schottky anomaly (C\symbol{126}%
1/T$^{2}$ above the peak) but rather seems to be a field-induced anomaly.
\ (The upturns in C/T for H $\geq$ 6 T are caused by the applied field
splitting the nuclear magnetic moment energy levels and creating a Schottky
peak in the specific heat.)%

%TCIMACRO{\FRAME{ftbpFU}{3.2897in}{2.3774in}{0pt}{\Qcb{Field-induced anomaly in
%C/T for x=0.95}}{}{lafig6js7.eps}{\special{ language "Scientific Word";
%type "GRAPHIC";  maintain-aspect-ratio TRUE;  display "USEDEF";
%valid_file "F";  width 3.2897in;  height 2.3774in;  depth 0pt;
%original-width 3.5682in;  original-height 2.9006in;  cropleft "0";
%croptop "0.9242";  cropright "1";  cropbottom "0.0378";
%filename 'LAFIG6JS7.EPS';file-properties "XNPEU";}} }%
%BeginExpansion
\begin{figure}
[ptb]
\begin{center}
\includegraphics[
trim=0.000000in 0.109643in 0.000000in 0.219865in,
natheight=2.900600in,
natwidth=3.568200in,
height=2.3774in,
width=3.2897in
]%
{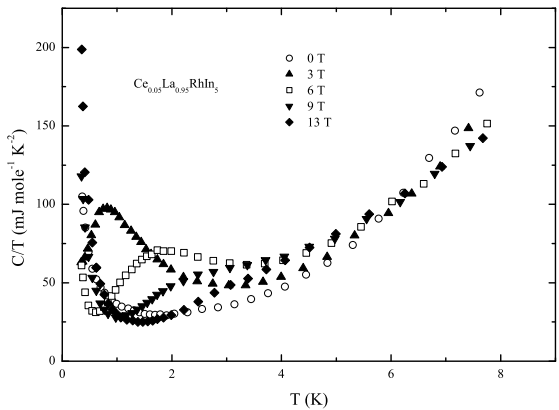}%
\caption{Field-induced anomaly in C/T for x=0.95}%
\end{center}
\end{figure}
%EndExpansion

\qquad Castro Neto and Jones have recently published$^{14}$ a theory of how
the specific heat and magnetization of materials with non-Fermi liquid
behavior caused by disorder-induced Griffiths phase spin clusters should scale
with magnetic field. \ In general, both the magnetization and specific heat
are predicted to exhibit low field behaviors (M \symbol{126}H and C/T
\symbol{126}T$^{-1+\lambda}$) which crossover over to the respective high
field behaviors (M \symbol{126}H$^{\lambda}$ and C/T \symbol{126}%
(H$^{2+\lambda/2}$/T$^{3-\lambda/2}$)e$^{-\mu_{eff}H/T}$) at the \textit{same}
magnetic field. \ The prediction for the field and temperature dependence for
the high field specific heat leads to a peak in C/T (or a shoulder in C) as a
function of increasing temperature - thus qualitatively consistent with the
data shown in Fig. 6. \ 

\qquad Although the specific heat data in field was taken in fairly widely
spaced fields, the fact that a peak occurs already in C/T in H=3 T offers a
prediction (the equality of the crossover field requires that the crossover
field for the magnetization data be perforce below 3 T) that can be checked by
examining the M vs H data, where a much more finely spaced sequence of fields
was used. \ In addition, the high field prediction that M \symbol{126}%
H$^{\lambda}$ can be checked up to 5.5 T, and this field-dependence
determination of $\lambda$ can then be compared with that independently
determined from the \textit{temperature dependence} of $\chi$ in Figure 4.
\ Thus, magnetization data for both field directions for single crystal
Ce$_{0.05}$La$_{0.95}$RhIn$_{5}$ are shown fitted to these Griffiths phase low
and high field predictions in Figures 7 and 8, H $\parallel$,$\perp$ basal
plane respectively. \ As may be seen, using the values for $\lambda_{\chi}$
determined from Fig. 4 (0.14 and 0.41 for H($\parallel$,$\perp$) basal plane
respectively) gives rather good$^{20}$ agreement between the predicted, M
\symbol{126}H$^{\lambda}$ dependence and the high field magnetization data.
\ \ (The fit to the higher field data with the lowest standard deviation
actually gives $\lambda$=0.67; however, the standard deviations are within 8\%
of one another.) \ Further, the deviation from linear behavior at low fields
occurs (see Figs. 7 and 8) above 0.8 T and the deviation from the M
\symbol{126}H$^{\lambda}$ power law occurs below 1.2 T. \ These estimates for
the crossover field are not inconsistent with the peak in C/T (where a peak is
characteristic of the high field regime) occuring in 3 T, Fig. 6. \ (Work
under way$^{21}$ to more thoroughly characterize the low and high field
behavior for M and C/T for x=0.95 has found that a peak in C/T field data
taken in 0.5 T increments down to 0.3 K first appears at 1.5 T.)%

%TCIMACRO{\FRAME{ftbpFU}{3.2889in}{2.4491in}{0pt}{\Qcb{M vs H for x=0.95, H in
%basal plane}}{}{lafig7js7.eps}{\special{ language "Scientific Word";
%type "GRAPHIC";  maintain-aspect-ratio TRUE;  display "USEDEF";
%valid_file "F";  width 3.2889in;  height 2.4491in;  depth 0pt;
%original-width 3.5561in;  original-height 2.9672in;  cropleft "0";
%croptop "0.9262";  cropright "1";  cropbottom "0.0368";
%filename 'LAFIG7JS7.EPS';file-properties "XNPEU";}} }%
%BeginExpansion
\begin{figure}
[ptb]
\begin{center}
\includegraphics[
trim=0.000000in 0.109193in 0.000000in 0.218979in,
natheight=2.967200in,
natwidth=3.556100in,
height=2.4491in,
width=3.2889in
]%
{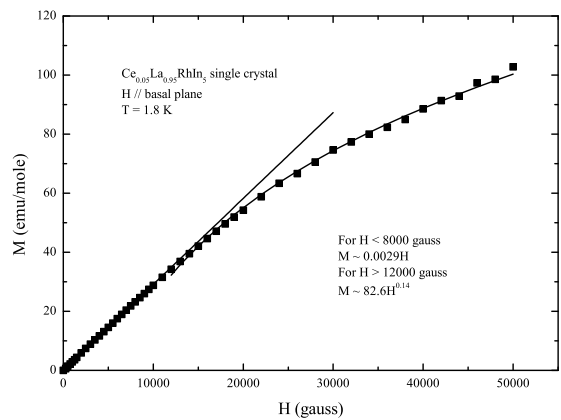}%
\caption{M vs H for x=0.95, H in basal plane}%
\end{center}
\end{figure}
%EndExpansion

\qquad%
%TCIMACRO{\FRAME{ftbpFU}{3.2889in}{2.4362in}{0pt}{\Qcb{M vs H for x=0.95, H
%perp. to basal plane}}{}{lafig8js7.eps}{\special{ language "Scientific Word";
%type "GRAPHIC";  maintain-aspect-ratio TRUE;  display "USEDEF";
%valid_file "F";  width 3.2889in;  height 2.4362in;  depth 0pt;
%original-width 3.5561in;  original-height 2.9551in;  cropleft "0";
%croptop "0.9259";  cropright "1";  cropbottom "0.0370";
%filename 'LAFIG8JS7.EPS';file-properties "XNPEU";}} }%
%BeginExpansion
\begin{figure}
[ptb]
\begin{center}
\includegraphics[
trim=0.000000in 0.109339in 0.000000in 0.218973in,
natheight=2.955100in,
natwidth=3.556100in,
height=2.4362in,
width=3.2889in
]%
{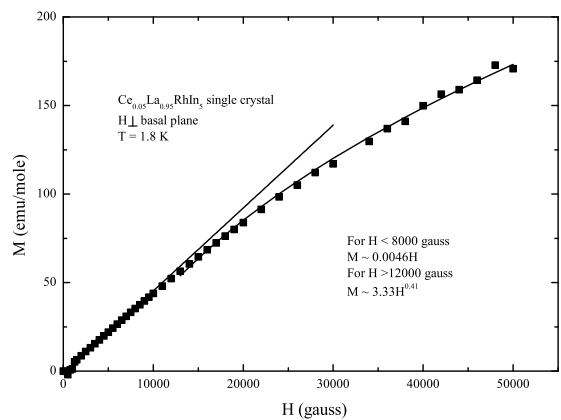}%
\caption{M vs H for x=0.95, H perp. to basal plane}%
\end{center}
\end{figure}
%EndExpansion

\qquad Another prediction$^{14}$ of the Griffiths phase theory of Castro Neto
and Jones, the field and temperature dependence of C/T in the high field
limit, is compared$^{22}$ to the 3T Ce$_{0.05}$La$_{0.95}$RhIn$_{5}$\ data
(with the fit$^{18}$ to pure LaRhIn$_{5}$ and the small,
%TCIMACRO{\TEXTsymbol{<}}%
%BeginExpansion
$<$%
%EndExpansion
10\% at the lowest temperature, contribution due to the field splitting of the
nuclear moments, subtracted off), H $\parallel$ basal plane, in Fig. 9.
\ \ Using only two fit parameters (the amplitude and the effective moment,
$\mu_{eff}$) and fixing $\lambda$ = 0.14 (based on $\lambda_{\chi}$) gives the
fit (dashed line in Fig. 9) as shown, with the reasonable$^{14,23}$ fitted
value for $\mu_{eff}$\ (which corresponds to the average moment in the
Griffiths phase spin cluster) of 1.25 $\mu_{B}$. \ \ Clearly, fitting C/T to
(H$^{2+\lambda/2}$/T$^{3-\lambda/2}$)e$^{-\mu_{eff}H/T}$ is a fairly good
representation of the data. \ (To give an idea how the fit depends on the
effective moment, a fit to these 3 T data with $\mu_{eff}$ constrained to be
1.0 $\mu_{B}$ is shifted by to lower temperatures by \symbol{126}0.2 K from
the present fit.)\ \ %

%TCIMACRO{\FRAME{ftbpFU}{3.2897in}{2.3272in}{0pt}{\Qcb{Fit of field-induced
%anomaly to theory}}{}{lafig9js7.eps}{\special{ language "Scientific Word";
%type "GRAPHIC";  maintain-aspect-ratio TRUE;  display "USEDEF";
%valid_file "F";  width 3.2897in;  height 2.3272in;  depth 0pt;
%original-width 3.6391in;  original-height 2.9006in;  cropleft "0";
%croptop "0.9228";  cropright "1";  cropbottom "0.0385";
%filename 'LAFIG9JS7.EPS';file-properties "XNPEU";}} }%
%BeginExpansion
\begin{figure}
[ptb]
\begin{center}
\includegraphics[
trim=0.000000in 0.111673in 0.000000in 0.223926in,
natheight=2.900600in,
natwidth=3.639100in,
height=2.3272in,
width=3.2897in
]%
{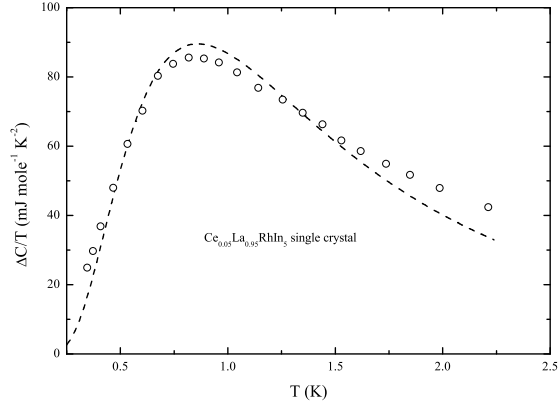}%
\caption{Fit of field-induced anomaly to theory}%
\end{center}
\end{figure}
%EndExpansion

\textbf{IV \ Conclusions}

\qquad Despite the difficulty of precisely compensating for the broad peak in
C/T in Ce$_{1-x}$La$_{x}$RhIn$_{5}$ centered at about 3 K, the apparent
$\gamma$ per Ce mole for x$\geq$0.5, away from the antiferromagnetic
transition in the phase diagram, appears to be less than 100 mJ/Ce-moleK$^{2}$
- in disagreement with estimates for $\gamma$ in the literature$^{4-5}$ but
not inconsistent with the dHvA results of Alvers et al.$^{7}$ \ There is a
strong upturn in C/T below 1 K for x$\geq$0.5 that, when compared to the
temperature dependence of the susceptibility and the non-linear M vs H data,
is consistent with non-Fermi liquid behavior due to disordered spin clusters
('Griffiths phases.') \ \ Applied magnetic field suppresses this upturn in C/T
already by 3 T; above 3 T the C/T results show a broad anomaly that further
broadens and moves to higher temperatures as field is increased. \ This field
induced anomaly, together with the field dependence of the magnetization,
compares well with the predictions of the Griffiths phase theory$^{14,24}$ of
Castro Neto and Jones, particularly in the magnetization data as a function of
field and the agreement of these data with the predicted $\lambda_{\chi}$
exponent from the temperature dependence of the susceptibility. \ In summary,
the breadth of behavior observed in Ce$_{1-x}$La$_{x}$RhIn$_{5}$ in zero and
applied field is indicative of a phase diagram of unusual richness and variety.

\bigskip

\textbf{Acknowledgements: \ }The authors wish to thank Antonio Castro Neto for
quite fruitful discussions. \ \ Work at the University of Florida by performed
under the auspices of the United States Department of Energy, contract no.
DE-FG05-86ER45268. \ Partial summer support for J. Alwood and P. Watts from
the NHMFL and University of Florida NSF REU programs respectively is
gratefully acknowledged.

\qquad\qquad\qquad\qquad\qquad\qquad\qquad\qquad\qquad\qquad\qquad\qquad References

\bigskip

1. C. Petrovic, R. Movshovich, M. Jaime, P. G. Pagliuso, M. F. Hundley, J. L.
Sarrao, Z. Fisk, and J. D. Thompson, Europhys. Lett. \textbf{53}, 354 (2001).

2. C. Petrovic, P. G. Pagliuso, M. F. Hundley, R. Movshovich, J. L. Sarrao, J.
D. Thompson, and Z. Fisk, J. Condens. Mat. Phys. \textbf{13}, L337 (2001).

3. N. J. Curro, P. C. Hammel, P. G. Pagliuso, J. L. Sarrao, J. D. Thompson,
and Z. Fisk, Phys. Rev. B 62, 6100 (2000); W. Bao, P. G. Pagliuso, J. L.
Sarrao, J. D. Thompson, Z. Fisk, J. W. Lynn, and R. W. Irwin, Phys. Rev. B 62,
14621 (2000); W. Bao, P. G. Pagliuso, J. L. Sarrao, J. D. Thompson, Z. Fisk,
J. W. Lynn, and R. W. Irwin, Phys. Rev. B \textbf{63}, 219901(E) (2001).

4. H. Hegger, C. Petrovic, E. G. Moshopoulou, M. F. Hundley, J. L. Sarrao, Z.
Fisk, and J. D. Thompson, Phys. Rev. Lett. \textbf{84}, 4986 (2000).

5. \ A. L. Cornelius, P. G. Pagliuso, M. F. Hundley, and J. L. Sarrao, Phys.
Rev. B \textbf{64}, 144411 (2001).

6. \ J. S. Kim, J. Alwood, G. R. Stewart, J. L. Sarrao, and J. D. Thompson,
Phys. Rev. B \textbf{64}, 134524 (2001).

7. \ U. Alver, R. G. Goodrich, N. Harrison, D. W. Hall, E. C. Palm, T. P.
Murphy, S. W. Tozer, P. G. Pagliuso, N. O. Moreno, J. L. Sarrao, and Z. Fisk,
Phys. Rev. B \textbf{64}, 180402R (2001).

8. \ B. Andraka, C.S. Jee, J.S. Kim, Hauli Li, M.W. Meisel, and G.R. Stewart,
Physica B\textbf{171}, 384 (1991).

9. \ C.S. Jee, B. Andraka, J.S. Kim, Hauli Li, M.W. Meisel, and G.R. Stewart,
Phys. Rev. B\textbf{42}, 8630 (1990); J. S. Kim, C. S. Jee, W. W. Kim, B.
Andraka, P. Kumar, and G. R. Stewart, Phys. Rev. B\textbf{44}, 7473 (1991).

10. \ J. S. Kim, B. Andraka, and G. R. Stewart, Phys. Rev. B\textbf{44}, 6921 (1991).

11. \ G. R. Stewart, Rev. Sci. Instrum. \textbf{54}, 1 (1983).

12. \ For a review of non-Fermi liquid behavior, see G. R. Stewart, Rev. Mod.
Phys. \textbf{73}, 797 (2001).

13. \ A. H. Castro Neto,G. Castilla, and B. A. Jones, Phys. Rev. Lett.
\textbf{81}, 3531 (1998).

\bigskip14. \ A. H. Castro Neto and B. A. Jones, Phys. Rev. B \textbf{62},
14975 (2000).

15. \ M. C. DeAndrade, R. Chau, R. P. Dickey, N. R. Dilley, E. J. Freeman, D.
A. Gajewski,

M. B. Maple, R. Movshovich, A. H. Castro Neto, G. Castilla, and B. A. Jones,

Phys. Rev. Lett. \textbf{81}, 5620 (1998).

16. \ G. R. Stewart, Rev. Mod. Phys. \textbf{56}, 755 (1984).

17. \ For a discussion of spin cluster, spin glass, and Griffiths phase
behavior, see ref. 12.

18. \ Mike Hundley, private communication. \ See also ref. 4.

19. \ Note of course the unavoidable uncertainty is fitting the 'hump' - which
may very well involve entropy due to Ce-Ce interactions - to data from a more
dilute composition and then applying this fit to more concentrated systems.

20. \ The ''best fit'' value for the exponent $\lambda$ from the field
dependence of the magnetization for H $\parallel$ basal plane shown in Fig. 7
is within 0.1 of the value $\lambda_{\chi}$ = 0.14 determined from the
temperature dependence of $\chi$ determined in Fig. 4, i. e. within the error
bar. \ For H $\perp$ basal plane, the best fit to the magnetization data shown
in Fig. 8 gives $\lambda$ = 0.67 instead of the value determined from the
temperature dependence of $\chi$, where $\lambda_{\chi}$ = 0.41. \ However,
the standard deviation for the fit (to 20 data points) using $\lambda_{\chi}$
= 0.41 is less than 8\% higher than that for the ''best'' fit.

21. \ J. S. Kim, J. Alwood, D. Mixson, and G. R. Stewart, to be published.

22. \ Fits to the \ 6 and 9 T data are similar, although the correction for
the low temperature upturn in C/T caused by the nuclear hyperfine level
splitting due to the applied field is larger and the size of the field-induced
anomaly in C/T with increasing field is rapidly decreasing. \ Since the
crossover field between low and high field dependences, as determined by the
magnetization, is \symbol{126} 0.8 -1.2 T, the 3 T data should be well in the
high field limit.

23. \ A. H. Castro Neto, private communication.

24. \ Although a recent paper (A. J. Millis, D. K. Morr, and J. Schmalian,
Phys. Rev. Lett. \textbf{87}, 167202 \{2001\}) has called the theory of Castro
Neto and Jones into question based on dissipation arguments in the single
impurity limit, an even more recent work by Castro Neto and Jones
(cond-mat/0106176) argues that for concentrated systems the results of ref. 14
still hold.
\end{document}